\documentstyle[prb,multicol,aps,epsf]{revtex}
\begin{document}
\draft
\title{Anisotropic in-plane thermal conductivity of single-crystal 
YBa$_2$Cu$_4$O$_8$}
\author{J.~L.~Cohn}
\address{Physics Department, University of Miami, Coral Gables, FL 33124}
\author{J.~Karpinski}
\address{Laboratorium f\"ur Festk\"orperphysik, Eidgen\"ossiche, Technische 
Hochschule Z\"urich, CH-8093 Z\"urich, Switzerland}

\maketitle

\begin{abstract}
We report measurements of the in-plane thermal conductivity ($\kappa$) 
of YBa$_2$Cu$_4$O$_8$ (Y-124)
single crystals in the temperature range $4K\leq T\leq 300K$ and
compare them with previous results on YBa$_2$Cu$_3$O$_{6+x}$ (Y-123).  
For transport 
perpendicular to the CuO chains, $\kappa_a(300K)\simeq 10$W/mK, and 
along the chains, $\kappa_b(300K)\simeq 40$W/mK.  The temperature 
dependence of $\kappa$ for both transport directions is much stronger than
in YBa$_2$Cu$_3$O$_{6+x}$ (Y-123), indicative of substantially 
superior lattice heat conduction in Y-124,
and resulting in maximum values for $\kappa$ at $T\sim 20$K exceeding 200W/mK. 
The data imply a surprisingly large anisotropy in the lattice conduction. 
$\kappa_a$ and $\kappa_b$ are enhanced in the superconducting state as in 
other cuprates. The magnitude and anisotropy of the 
enhancement are discussed and compared to those of Y-123.  
\end{abstract}


\begin{multicols}{2}
\narrowtext
 
The in-plane thermal conductivity ($\kappa_{ab}$) of cuprate 
superconductors 
has been studied extensively over the past several years.\cite{UherRev}  
Unresolved issues remain concerning the normal-state temperature 
dependence and the superconducting-state enhancement.
Because of the low carrier density,  
heat conduction by the lattice accounts for more than half of the
measured normal-state $\kappa_{ab}$ in these materials.  One would 
expect $\kappa_{ab}(T)$ in high-quality crystals to reflect
a dominant lattice contribution similar to that of crystalline insulators, 
i.e. $\kappa_{ab}\sim 1/T$ at high-$T$, rising sharply to a maximum at low $T$.
This behavior is observed in Nd$_{2-x}$Ce$_x$CuO$_4$(Nd-214),\cite{CohnNCCO}
but for all other cuprates investigated $\kappa_{ab}$ is very weakly 
$T$ dependent for $T>T_c$.  Whereas for Bi$_2$Sr$_2$CaCuO$_8$ structural
disorder (e.g., the Bi-O layer modulation\cite{Chen}) might explain its 
nearly glass-like lattice contribution,\cite{Allen} this would not seem 
applicable to YBa$_2$Cu$_3$O$_{6+x}$ (Y-123) and La$_{2-x}$Sr$_x$CuO$_4$ 
(La-214).  Furthermore, $\kappa_{ab}(T)$ for undoped, insulating Y-123 and 
La-214 
is unconventional,\cite{CohnInsulators} suggesting that 
the weak $T$-dependence of $\kappa_{ab}$ is generic to the latter 
materials and not directly related to the presence of free charge.  
An unidentified phonon damping mechanism appears to
underlie this behavior, possibly due to local lattice 
distortions\cite{CohnInsulators} and/or magnetic excitations.\cite{Inyushkin}
Regarding the enhancement,\cite{Allen,enhancement}  thermal Hall conductivity 
experiments\cite{Krishana} on optimally-doped Y-123 imply that the phenomenon 
is largely electronic in origin, but less is known about underdoped compounds.
The magnitude of the enhancement correlates with 
the superconducting pair density\cite{CohnCorr,Popoviciu} in 
Y-123 throughout the underdoped regime,
and it is of interest to explore the generality 
of this behavior through studies of other cuprates.

Here we report measurements of the in-plane thermal conductivity of 
YBa$_2$Cu$_4$O$_8$ (Y-124).  This material is of particular interest with 
regard to the issues mentioned above because of its structural 
similarity to Y-123, absence of oxygen vacancies on the CuO chains, and
the naturally underdoped state of the CuO$_2$ planes. 
We find that the magnitude of $\kappa_a$ (transverse to the chains) at 
$T$=300K is comparable to that of Y-123, but $\kappa_b$ is 3-4 times larger.
A very large in-plane anisotropy in the lattice conduction is implied.  
Both $\kappa_a(T)$ and $\kappa_b(T)$ behave as in conventional crystalline 
insulators (like Nd-214); their maximum values
(at $T\simeq 20$K) are the highest reported for any cuprate and exceed 
those of Y-123 by an order of magnitude.  These results imply 
a strong damping of phonons by static or dynamic structural distortions
as the source for the much weaker $T$ dependence of $\kappa_{ab}$ in Y-123. 
The superconducting-state enhancement for Y-124 is comparable in magnitude 
to that of $T_c$=60K Y-123, consistent with the underdoped state of the planes 
in the double-chain compound.  We discuss the anisotropy of the enhancement and 
its implications for the Lorenz number.

Thermal conductivity measurements were performed on three single crystals 
grown by a high-pressure flux method as described 
previously.\cite{CrystalGrowth}  
Two of these (189 and 259) were grown in Al$_2$O$_3$ crucibles
yielding a slight Al contamination, and $T_c=72$K.  Recent analyses
indicate\cite{Alsubstitution} that Al substitutes for 1-2\%
of the Cu(2) atoms in the CuO$_2$ planes.  As we discuss below, this light 
doping has a substantial effect on the thermal conductivity.
The third crystal (315), grown in Y$_2$O$_3$ and without Al 
contamination, had $T_c=80$K.  
Typical crystal dimensions were $0.8\times 0.5\times0.05$mm$^3$,
with the shortest dimension along the crystalline $c$ axis.  The $a$- 
and $b$-axis electrical resistivities of similarly-prepared crystals have 
been discussed extensively elsewhere;\cite{Rhopapers} typical values at 
$T=$300K are $\rho_a=400\mu\Omega$cm and $\rho_b=130\mu\Omega$cm. 
The steady-state thermal conductivity measurements
employed a fine-wire differential chromel-constantan thermocouple and 
miniature chip resistor as heater, both glued to the specimen with varnish 
or epoxy.  The absolute accuracy of the measurements is $\pm 20$\% due to 
uncertainty in the placement of the thermocouple junctions.   No 
corrections for heat losses (via radiation and conduction through the leads) 
have been applied; these are estimated to be $\sim 10$\% 
near room temperature and $\alt 2$\% at $T\alt 120$K.  For specimens 259 
and 315 
gold contacts were vapor deposited and their a-axis thermoelectric 
powers measured, yielding $S(290K)=36\mu$V/K and
16$\mu$V/K, respectively.  These values provide estimates of the hole 
concentration per planar Cu atom,\cite{TEPmeasuresp} $p\simeq 0.09$ 
and 0.11 for the Al-doped and undoped specimens, respectively.  

The results for $\kappa_a$ and $\kappa_b$ are shown in Fig.~\ref{KappavsT}.
The anisotropy in the normal state is substantial, with 
$\kappa_b/\kappa_a\sim 3-4$ (Fig.~\ref{Anisotropy}).
The normal-state $T$ dependence in both crystallographic directions is close to
$1/T$ for specimen 315 and is significantly weaker ($\sim 1/T^{1/2}$)
for the crystals containing Al.  In all cases the $T$-dependence is 
substantially stronger than that of both $T_c=92$K and 60K Y-123 
(solid lines, Fig.~\ref{KappavsT}).  
At low $T$ the characteristic dielectric maxima are observed near 20K.  
A change in slope is evident at $T_c$ 
[Fig.~\ref{KsKnslopes}~(a)] for all specimens, though not appearing as 
prominent as in Y-123 
because of the substantially stronger normal-state $T$ dependence.  The 
low-$T$ maximum values,
$\kappa_b\simeq 140$ and 245 W/mK for samples 189 and 259, 
respectively (and 
500W/mK for crystal 315 by extrapolation), are the 
largest reported for any cuprate, exceeding the 100W/mK of insulating 
Nd-214,\cite{CohnNCCO} and 25-40 W/mK observed in untwinned 
$T_c=92$K Y-123.\cite{Uher,Cohnuntwnd,Yu,Gold,GagnonK} These observations
indicate substantially superior and rather conventional 
lattice conduction in Y-124, confirming
a similar conclusion based on previous measurements of 
polycrystals.\cite{Andersson}
\vfill
\centerline{\epsfxsize=3.25in\epsfbox[50 70 540 730]{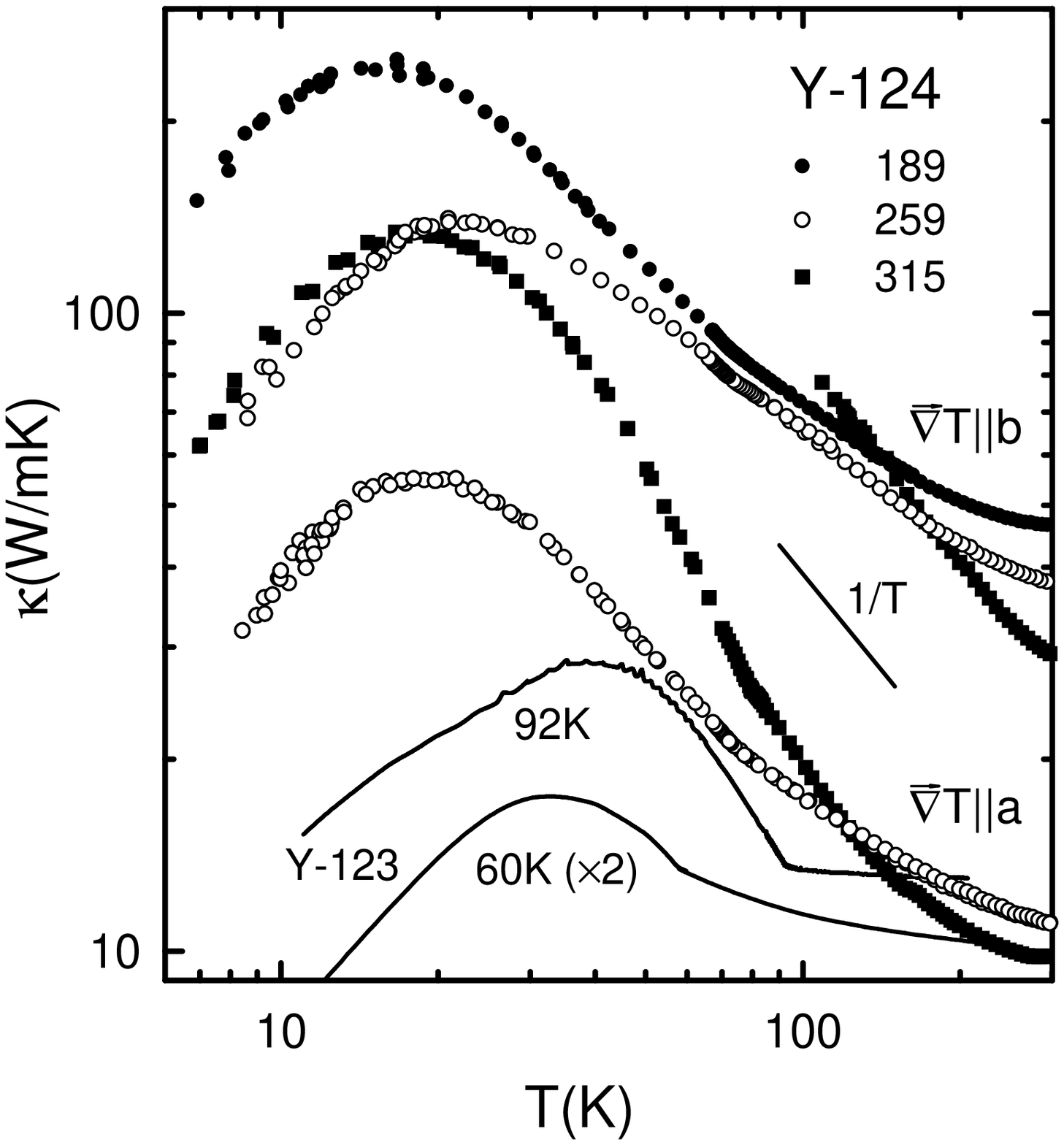}}
\vskip -1.35in
\begin{figure}
\caption{$\kappa_a(T)$ and $\kappa_b(T)$ for three Y-124 crystals. 
Specimens 189 and 259 are lightly Al-doped. Data for Y-123 crystals 
(solids lines) are from Ref.~\protect{\onlinecite{Cohnuntwnd}} ($T_c=92$K)
and Ref.~\protect{\onlinecite{Inyushkin}} ($T_c=60$K).} 
\label{KappavsT}
\end{figure}
The total thermal conductivity is a sum of electronic and lattice components,
$\kappa=\kappa_e+\kappa_L$.  First let us consider the electronic 
contributions, 
$\kappa_{a,e}$ and $\kappa_{b,e}$.  The 
Wiedemann-Franz law, $\kappa_e(T)=L_0T/\rho$ 
($L_0=2.44\times 10^{-8}$W$\Omega$/K$^2$), and electrical resistivities
\cite{Rhopapers} 
provide upper-bound estimates: $\kappa_{e,a}(300K)\approx 2$W/mK and
$\kappa_{e,b}(300K)\approx 6$W/mK.  These imply $\kappa_{L,a}\approx 8$W/mK
and $\kappa_{L,b}\approx 25-39$W/mK, and the very large in-plane 
lattice anisotropy, $\kappa_{L,b}/\kappa_{L,a}\sim 3-4$. 
This is to be contrasted with Y-123 where similar analyses suggest
that $\kappa_L$ is nearly isotropic in the planes.
\cite{Cohnuntwnd,Yu,Gold,GagnonK}
Since Y-124 
and Y-123 differ only in their CuO-chain structures, we must conclude that 
the strong lattice anisotropy in Y-124 is directly 
connected with the chain-related vibrational spectrum.

Anisotropy in $\kappa_L$
can arise from either anisotropy of the phonon group velocities, their
relaxation times, or both.  Though optic modes can contribute to heat 
transport and the anisotropy, their contribution should diminish 
with decreasing temperature as should any associated anisotropy.\cite{Jaffe}
This is
contrary to our observation that $\kappa_b/\kappa_a$ in the normal-state
is nearly constant or increases with decreasing $T$ (Fig.~\ref{Anisotropy}) 
[the decrease in $\kappa_b/\kappa_a$ below $T_c$ is related to 
superconductivity, as discussed below]. Experimental dispersion curves are 
not available for Y-124, however the computed spectrum\cite{Y124phonons} 
is quite similar to that of Y-123, and does not imply a substantial
in-plane anisotropy in the acoustic-mode group velocities.  These observations 
would suggest a large anisotropy in the phonon relaxation rates in Y-124, but
further information about the low-energy vibrational spectrum is clearly 
needed to address this issue.

The rather strong influence of the light Al doping in the planes on $\kappa$
provides insight into the phonon
scattering in both Y-124 and Y-123.  Nuclear quadrupole\cite{Alsubstitution}\break
\vfill
\centerline{\epsfxsize=3.25in\epsfbox[50 60 540 570]{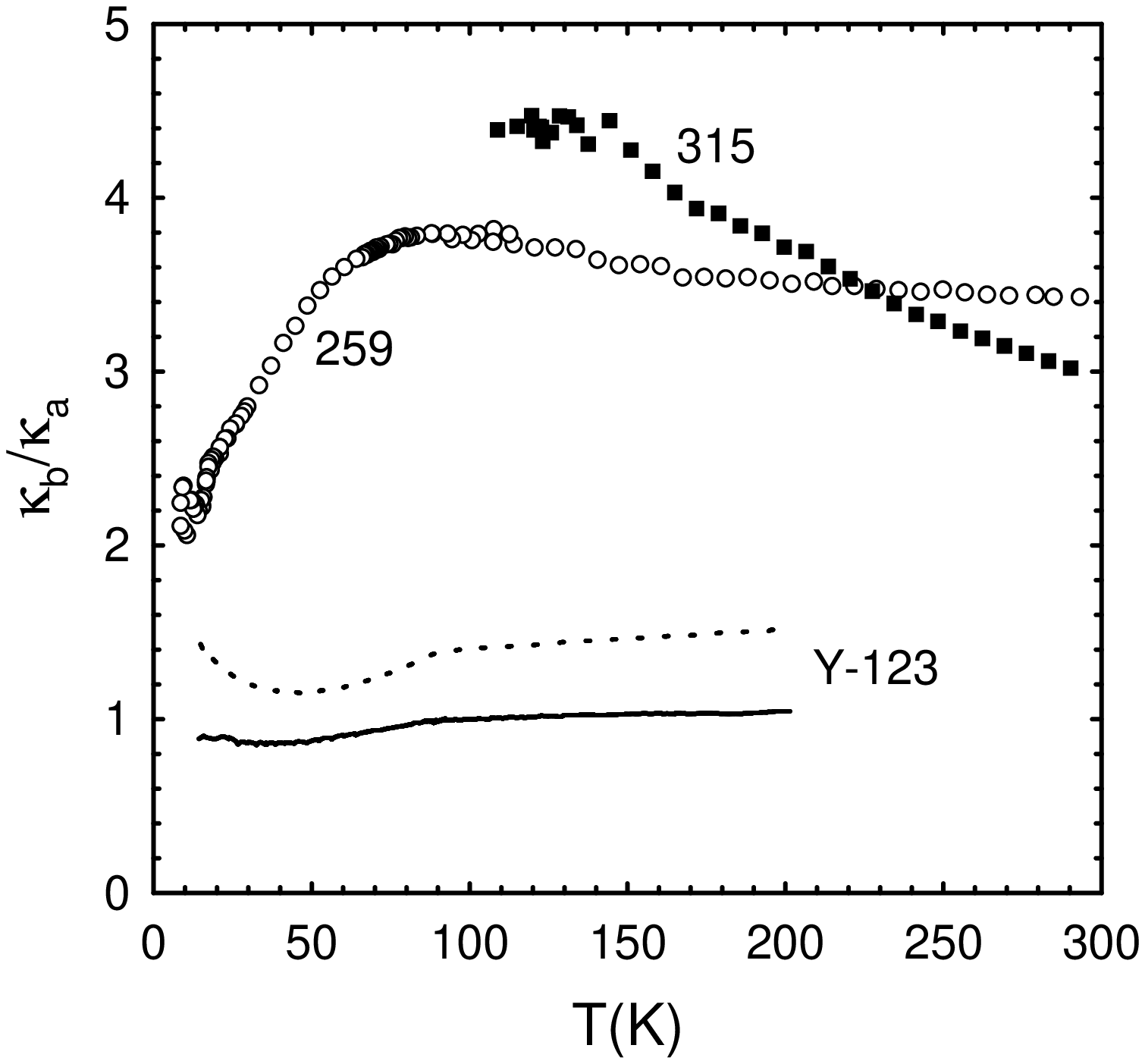}}
\vskip -2in
\begin{figure}
\caption{Anisotropy, $\kappa_b/\kappa_a$ {\it vs} $T$ for Y-124 crystals. 
Also shown are data for untwinned $T_c=92$K Y-123 from 
Ref.~\protect{\onlinecite{Cohnuntwnd}} (solid curve) and
Ref.~\protect{\onlinecite{Yu}} (dashed curve).}
\label{Anisotropy}
\end{figure}
\noindent
resonance studies of similar Y-124 crystalsreveal substantially 
larger $^{63}$Cu linewidths in the Al-doped crystals,
attributed to static 
disorder in 
the Cu(2)-apical oxygen bondlengths (a typical Al-O bondlength for octahedral 
coordination
is 1.94$\rm \AA$ as compared with 2.28$\rm \AA$ for Cu-O).  Such local 
structural distortions evidently represent a much more effective phonon 
scattering 
mechanism than does the mass defect\cite{massdefect,Berman} introduced by the 
substitution.  
The local structural modifications may also explain why the hole concentration 
and $T_c$ are so 
dramatically suppressed -- the 2\% Al substitution for planar Cu
reduces the mobile planar hole concentration by about 20\%, i.e. a ten-fold 
larger reduction than would be expected if each Al atom filled one hole.
This observation suggests that the suppression of mobile holes, by the 
combined effects of charge compensation, disorder-induced localization, and 
changes in the local electronic structure, extends to next nearest neighbors 
of each Al atom in the planes.

In Y-123 a similar static disorder in the apical bondlengths is induced by 
oxygen vacancies on the chains\cite{Jorgensen} and there is evidence
for static or dynamical structural distortions of the CuO$_2$ 
planes in the normal state,\cite{distortions} possibly related to 
oxygen vacancies or diffusion that are absent in Y-124.  
The implication is that these features
give rise to substantial damping of in-plane heat-carrying phonons, and are 
the origin of the weaker 
$T$-dependence and in-plane anisotropy observed for $\kappa$ in the 
single-chain compound.

It is of interest to compare the superconducting-state enhancement of $\kappa$
for Y-124 with that of Y-123.\cite{CohnCorr,Popoviciu}  We define the 
enhancement as\cite{Popoviciu} 
$\Gamma\equiv -d(\kappa^s/\kappa^n)/dt|_{t\to 1}$, 
the reduced temperature derivative
of the superconducting to normal-state thermal conductivity ratio near 
$T_c$.  $\kappa^n$ is determined by a 
polynomial extrapolation of 
the normal-state data just below $T_c$ as shown in Fig.~\ref{KsKnslopes}~(a).
Figure~\ref{KsKnslopes}~(b) shows the normalized data
and corresponding slopes.

The $\Gamma$ values agree with those found for
twinned Y-123 at similar $p$ values,\cite{Popoviciu} 
indicating that for the regime 
investigated here ($p\leq 0.11$), $\Gamma$ is largely determined by the
change in scattering that occurs in the CuO$_2$ planes. 
That $\Gamma_a$ is larger for Al-free specimen 315 than for specimen 259 
is consistent with the higher mobile hole concentration in the planes of the 
former.  Also of interest is the anisotropy: $\Gamma_a/\Gamma_b\approx 1.5$ for 
specimen 259.  Evidently this anisotropy 
is responsible for the 
decrease in $\kappa_b/\kappa_a$ at $T\leq T_c$ (Fig.~\ref{Anisotropy}).  
A similar 
behavior was observed for untwinned Y-123 ($x\sim 0.9$)
[solid and dashed lines in Fig.~\ref{Anisotropy}], where the enhancement is 
also anisotropic,\cite{Cohnuntwnd,Yu,Gold,GagnonK} with a slightly smaller 
$\Gamma_a/\Gamma_b\approx 1.2-1.5$.

Consider $\Gamma$ in more detail.  We may write,
\begin{equation}
\Gamma_i=(\kappa_{e,i}/\kappa_i)\Gamma_{e,i}+
(\kappa_{L,i}/\kappa_i)\Gamma_{L,i},\label{Gamfull}
\end{equation}
where $i=a,b$, and $\Gamma_{e,i}$ 
and $\Gamma_{L,i}$ are the $i$-axis electronic and lattice slope changes,
respectively.  The thermal conductivities are evaluated at $T_c$.  
Theoretically\cite{AmbegaokarWoo}
$\Gamma_e$ is a\break 
\vglue 1.3in
\centerline{\epsfxsize=3.25in\epsfbox[50 60 540 570]{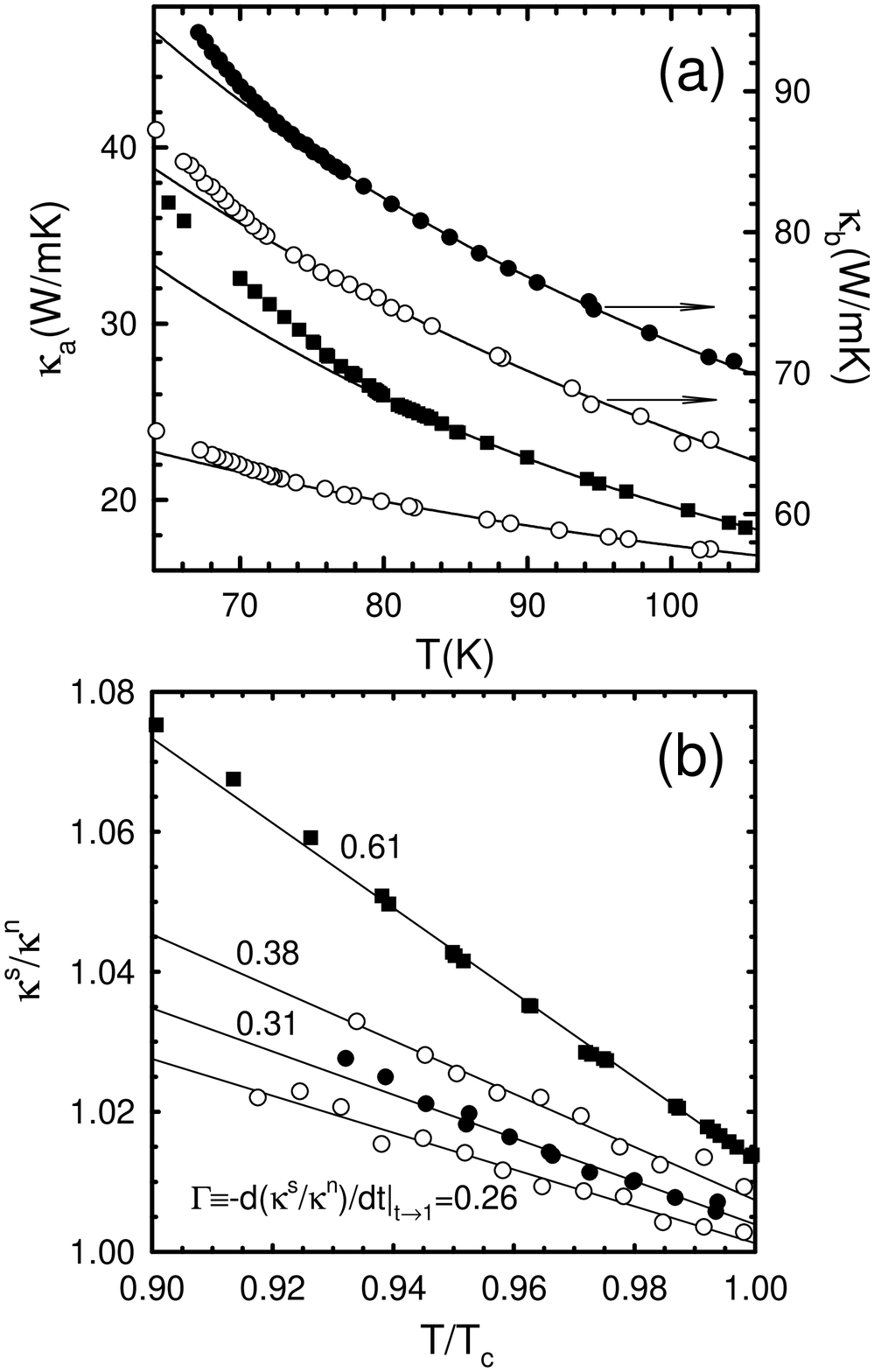}}
\vskip .2in
\begin{figure}
\caption{(a) $\kappa_a(T)$ and $\kappa_b(T)$ showing the slope change
near $T_c$ on an expanded scale. Specimens are represented by the 
same symbols as in Fig.~1.  Solid lines are polynomial 
fits to the normal-state data, defining $\kappa^n$. (b) ratio of 
superconducting to normal-state thermal conductivity {\it vs} reduced 
temperature.  Solid lines are linear-least-squares fits to the data
in the range $0.92\leq t\leq 1$, defining 
$\Gamma\equiv -d(\kappa^s/\kappa^n)/dt|_{t\to 1}$. 
The data for specimen 315 (solid squares) are shifted vertically by 0.01 
for clarity.} 
\label{KsKnslopes}
\end{figure}
\noindent
sum of two terms: one proportional to $d\Delta^2/dt|_{t=1}$
and the other proportional to $-d(\tau^s_{qp}/\tau_{qp}^n)/dt|_{t=1}$ 
($\Delta$ and $\tau_{qp}$ are the real part of the superconducting gap 
and quasiparticle (qp) lifetime, respectively).  In the cuprates, 
the second term 
predominates due to the strong suppression of qp 
scattering\cite{qpscattering} 
at $T<T_c$, and $\Gamma_e$ is positive. Scattering of phonons by charge 
carriers always produces a positive lattice term ($\Gamma_L$), but the 
thermal Hall conductivity measurements of Krishana {\it et al.}\cite{Krishana} 
indicate that the electronic contribution to the enhancement predominates in
optimally-doped Y-123.  
Their data for twinned Y-123 ($x\sim 0.9$) yields: 
$\Gamma_{ab}\simeq 1.4$, $\Gamma_{e,ab}\simeq 12.5$, and
$\kappa_{e,ab}/\kappa_{ab}\simeq 9\times10^{-2}$. Thus, 
$\Gamma_{L,ab}\simeq 0.3$, and about 80\% of the slope change for 
optimally-doped Y-123 is electronic in origin.
Recent measurements\cite{Krishanapreprint} suggest that the slope change in
underdoped Y-123 is also largely electronic in origin, and thus it is likely 
that this conclusion also applies to Y-124. 

Given these observations, one might expect that the $\Gamma$ anisotropy simply 
follows that of the superfluid, with the larger  qp lifetime enhancement 
occurring for transport along the crystallographic direction characterized by 
the largest superfluid density.  However, this is not the case; for Y-123 and 
Y-124 the superfluid density is 2-3 times larger along the 
chains,\cite{Zhang,Basov,TallonmuSR} whereas $\Gamma$ is larger along the 
planes.  To get further insight into the origin of the $\Gamma$ anisotropy, 
we examine the limiting case, $\Gamma_{L,i}=0$, for  
which $\Gamma_i$ is determined entirely by the electronic term.  The electronic 
superconducting-state slope change may be related to the 
low-frequency electrical conductivity via the Wiedemann-Franz 
law,\cite{Salamon} 
$\kappa_{e,i}=L_i\sigma_{1i}T$, where $L_i$ and 
$\sigma_{1i}$ are the Lorenz number and real part of the electrical 
conductivity, 
respectively.  Substituting into Eq.~\ref{Gamfull} we have
\begin{eqnarray}
\Gamma_i=-{L_i\sigma_{1i}T_c^2\over \kappa_i}\bigg[&&{1\over L_i}
\left({dL_i^s\over dT}-{dL_i^n\over dT}\right)\nonumber \\
&&+{1\over \sigma_1i}
\left({d\sigma_{1i}^s\over dT}-{d\sigma_{1i}^n\over dT} \right)
\bigg]_{T=T_c}
\label{Game}
\end{eqnarray}
\noindent
Theoretically, the first term has magnitude 
and sign that are highly model dependent.\cite{Graf}  
We first examine the second term.
For untwinned, 
optimally-doped Y-123,   
the microwave data of Zhang {\it et al.}\cite{Zhang} yield,
$(d\sigma_{1a}^s/dT)|_{T_c}\simeq -0.8\times10^5 (\Omega$ m K)$^{-1}$ and
$(d\sigma_{1b}^s/dT)|_{T_c}\simeq -1.8\times10^5 (\Omega$ m K)$^{-1}$.
For the normal-state data we employ $d\sigma/dT=-(1/\rho^2)d\rho/dT$
and an average of $\rho(T)$ measurements on high-quality untwinned 
crystals\cite{Rhountwnd}:
$\rho_a(T_c)\simeq 75\pm 25\mu\Omega$cm, 
$d\rho_a/dT|_{T_c}\simeq 0.85\pm 0.25\mu\Omega$cm/K,
$\rho_b(T_c)\simeq 38\pm 12\mu\Omega$cm, 
$d\rho_b/dT|_{T_c}\simeq 0.35\pm 0.10\mu\Omega$cm/K.
For the $\kappa$ data we use,\cite{Yu} $\kappa_a(T_c)=11.5$W/mK and 
$\kappa_b(T_c)=16.0$W/mK.  With these values, the second terms in the 
expressions for $\Gamma_a$ and 
$\Gamma_b$ are, $(1.2\pm 0.2)L_a/L_0$ and
$(2.0\pm 0.2)L_b/L_0$, respectively.   
Experiments\cite{Cohnuntwnd,Yu,Gold,GagnonK} 
indicate $\Gamma_a=1.5-1.6$ and $\Gamma_b=1.0-1.3$.  
One possibility is that the first terms, $(dL_i^s/dT-dL_i^n/dT)|_{T_c}$, are
negligible, whence the Lorenz numbers at $T_c$ are  
$L_a/L_0\simeq 1.3$ and $L_b/L_0\simeq 0.6$.  Though $L^n>L_0$  
has been predicted for strongly correlated 
systems,\cite{IoffeKotliar}
these values for $L_a$ and $L_b$ are incompatible with
the results of Ref.~\onlinecite{Krishana} which imply $L_{ab}\sim (0.3-0.5)L_0$
for twinned crystals.  The latter estimate 
is consistent with the theoretical analysis of Hirschfeld
and Putikka,\cite{enhancement} which incorporates inelastic spin-fluctuation 
scattering and parameters constrained to fit the normal-state NMR relaxation 
and electrical resistivity.  We thus conclude that both $L_a$ and $L_b$
are less than $L_0$ and that the $(dL_i^s/dT-dL_i^n/dT)|_{T_c}$ are 
negative and of comparable magnitude to the second terms in Eq.~\ref{Game}.
A considerable $a/b$ anisotropy ($\sim 2-3$) in the $L_i$, 
$(dL_i^s/dT-dL_i^n/dT)|_{T_c}$, or both is then implied.  Calculations 
which incorporate inelastic as well as elastic scattering to yield the 
$L_i(T)$ near $T_c$ would be useful to extract further information from 
the $\Gamma_i$.

Work at the University of Miami was supported, in part, by NSF Grant 
No.~DMR-9631236, and at the ETH by the Swiss National Science Foundation.

\end{multicols}

\begin{references}
\bibitem{massdefect} The scattering rate for mass defects can be expressed as 
(see Ref.~\protect{\onlinecite{Berman}}, p. 74),
$\tau^{-1}=(c_pa^3\omega^4/4\pi v^3)(\Delta M/M)^2$, where $c_p$ is the defect
concentration, $a^3$ the volume per defect, $v$ the sound velocity, $M$ the 
mass of the atom normally occupying the defect site, and $\Delta M$ the 
difference in mass represented by the defect.
Using $c_p=2.7\times10^{-3}$ (corresponding to 2\% Al doping for planar Cu), 
$a^3=1.1\times 10^{-29}m^3$, $v=2,900$m/s, and $\Delta M/M=0.2$ (we use the 
average mass per formula unit), yields 
$\tau^{-1}\simeq 4\times 10^{-45}\omega^4$ s$^{-1}$.  Calculations using 
the Debye model (Ref.~\protect{\onlinecite{Berman}}, p. 23)
indicate that this scattering rate is two orders of 
magnitude too small to account for the suppressed low-$T$ value of $\kappa_a$ 
in specimen 259 as compared with that of Al-free specimen 315.
\end{references}
\end{document}